\documentclass[12pt,superscriptaddress,times]{elsarticle}
\usepackage{times}

\makeatletter
\def\ps@pprintTitle{%
  \let\@oddhead\@empty
  \let\@evenhead\@empty
  \def\@oddfoot{\reset@font\hfil\thepage\hfil}
  \let\@evenfoot\@oddfoot
}
\makeatother

\usepackage{graphicx}
\usepackage{slashed}
\usepackage{amssymb}
\usepackage{amsmath}
\newcommand{\nc}{\newcommand}
\nc{\lb}{\langle}
\nc{\rk}{\rangle}
\nc{\Blb}{\Big\langle}
\nc{\Brk}{\Big\rangle}

\nc{\mi}{\!\!\mid\!\!}
\nc{\ra}{\rightarrow}
\nc{\Ra}{\Rightarrow}
\nc {\cd}{\partial}
\nc {\sla}{\slashed}
\nc{\ro}{\mathrm}
\nc{\ca}{\mathcal}
\nc{\bo}{\mathbf}
\nc{\Tr}{\ro{Tr}\,}
\nc{\Str}{\ro{Str}}
\nc{\realtrace}{\ro{Re\; Tr}}
\nc{\maxrealtrace}{\ro{max\, Re\; Tr}}
\nc{\ud}{\ro{d}}
\nc{\nn}{\nonumber}

\nc{\pb}{\bar{\psi}}
\nc{\p}{\psi}
\nc{\Pb}{\bar{\Psi}}
\nc{\vp}{\vec{\pi}}
\nc{\vap}{\varphi}
\nc{\vt}{\vec{\tau}}
\nc{\si}{\sigma}
\nc{\Si}{\Sigma}
\nc{\g}{\gamma}
\nc{\G}{\Gamma}
\nc{\la}{\lambda}
\nc{\La}{\Lambda}
\nc{\ep}{\epsilon}
\nc{\de}{\delta}
\nc{\De}{\Delta}
\nc{\cL}{\ca{L}}
\nc{\cLe}{\ca{L}_{\ro{eff}}}
\nc {\ti}{\tilde}
\nc{\f}{\frac}
\nc{\da}{\dagger}
\nc{\SU}{\ro{SU}}
\nc{\om}{\omega}
\nc{\Om}{\Omega}

\nc{\darrow}{\stackrel{\leftrightarrow}{\cd}}
\nc{\darrows}{\stackrel{\leftrightarrow}{\sla{\cd}}}
\nc{\Darrows}{\stackrel{\leftrightarrow}{\sla{D}}}
\nc{\mr}{\stackrel{\circ}{m}_\rho}

\nc{\rad}{\langle r^2\rangle}
\newcommand{\vk}{\vec{k}}
\newcommand{\vq}{\vec{q}}

\nc {\eqb}{\begin{equation}}
\nc {\eqe}{\end{equation}}
\nc {\eqab}{\begin{eqnarray}}
\nc {\eqae}{\end{eqnarray}}

\journal{Physics Letters B}

\begin{document}

\begin{frontmatter}

\title{Finite-volume corrections to charge radii}

\author[1]{J. M. M. Hall} 

\author[1]{D. B. Leinweber}

\author[1]{B. J. Owen}

\author[1,2]{R. D. Young} 
\address[1]{Special Research Centre for the Subatomic Structure of Matter 
  (CSSM), School of Chemistry and Physics, University of Adelaide 5005,
  Australia}
\address[2]{ARC Centre of Excellence for Particle Physics at the Terascale 
(CoEPP), School of Chemistry and Physics, University of Adelaide 5005,
  Australia}

%ELSARTICLE document class doesn't display preprint no.
%but for this article, it's:
%ADP-12-41/T808

%ABSTRACT
\begin{abstract}

The finite-volume nature of lattice QCD entails a variety of effects that 
must be handled in the process of performing chiral extrapolations. 
Since the pion cloud that surrounds hadrons becomes distorted 
in a finite volume, 
hadronic observables must be corrected  
before one can compare with the experimental values. 
The electric charge radius of the nucleon 
is of particular interest when considering 
the implementation of finite-volume corrections. 
It is common practice in the literature to transform electric 
form factors from the lattice into charge radii prior to analysis. 
However, there is a fundamental difficulty with using these charge radii 
 in a finite-volume extrapolation. The subtleties are a consequence 
of the absence of a  
continuous derivative on the lattice. 
A procedure is outlined for handling 
such finite-volume corrections, which must be applied directly to the 
electric form factors themselves rather than to the charge radii.

\end{abstract}


\end{frontmatter}

%*****************************************%
\section{Introduction}
\label{sect:intro}

Lattice QCD 
provides important non-perturbative techniques for 
the analysis of many observables. 
One of the notable features of lattice QCD is that it must be performed 
in a finite volume. The associated finite-volume effects can be used to access 
interesting phenomena. For example, 
multi-hadron states are only resolvable at finite lattice sizes; 
the discrete energy eigenvalues become increasingly close together 
as the box size becomes large.  
The finite-volume nature of lattice QCD
 has important consequences, some of which 
require careful attention. For example, although 
regularization in both the infrared and ultraviolet regions 
is an automatic feature of 
lattice QCD with a finite lattice spacing, 
finite-sized phenomena, such as the   
virtual pion clouds that surround hadrons, become distorted. 
This results in deviations in the values of lattice observables 
that can become significant in the chiral regime 
\cite{Gasser:1987zq,Leinweber:2003dg,Young:2004tb,Leinweber:2004tc,Beane:2004tw,Hall:2012pk}.  
Therefore, a method for correcting finite-volume effects by estimating their 
size is sought,
using a complementary approach, such as chiral effective field theory 
($\chi$EFT)
~\cite{Young:2004tb,Leinweber:2004tc,Beane:2004tw,Hall:2012pk,Young:2002cj,Leinweber:2006ug,Tiburzi:2007ep,Syritsyn:2009mx,Bratt:2010jn,Greil:2011aa,Hall:2010ai}.

Study of the quark mass dependence of lattice QCD simulation 
results can be particularly insightful for examining the 
chiral properties of hadrons. 
In relating lattice calculations to experimental results, 
it is essential to incorporate the 
low-energy features of QCD in order to obtain reliable extrapolations 
in both quark mass and volume. 

In lattice QCD, form factors are measured at discrete values of 
momentum transfer, 
corresponding to the quantization of the momentum modes on the finite 
spatial volume \cite{Syritsyn:2009mx,Bratt:2010jn,Leinweber:1990dv,Nozawa:1990gt,Leinweber:1992hy,Boinepalli:2006xd,Hedditch:2007ex,Boinepalli:2009sq,Yamazaki:2009zq,Alexandrou:2011db,Collins:2011mk}. 
Once form factors have been extracted from 
lattice simulations, they are typically converted directly 
into charge radii. 
The essential difficulty lies in the definition of the 
charge `radius' at finite volume (more precisely, the slope  
of the form factor at zero momentum transfer, $Q^2 = \vec{q}^2 - q_0^2$). 
In order to define the radius, a  
derivative must be applied to the electric form factor, 
with respect to a small momentum transfer. 
This approach breaks down on the lattice, where only discrete momentum values 
are allowed.

In most cases, calculating the finite-volume corrections poses 
no essential problems \cite{Beane:2004tw,Hall:2010ai,Hall:2012pk,Detmold:2004ap,Beane:2004rf,Young:2009zb,Hall:2011en}. 
However, because of the absence of a continuous derivative on the lattice, 
the treatment of the electric 
charge radius is more subtle \cite{Tiburzi:2007ep,Hu:2007eb,Jiang:2008ja}. 
Therefore, a method is outlined 
for handling finite-volume corrections to a given lattice simulation result. 

Finite-volume charge radii are calculated using the finite-volume electric 
form factors, $G_E^L(Q^2)$, with $Q^2$ taking an allowed value on the lattice. 
It will be shown that the finite-volume 
corrections to the loop integrals must be applied before the conversion 
from the form factor to the charge radius. 
An extrapolation in $Q^2$ 
is then chosen in order to construct an infinite-volume charge radius, 
defined in the usual manner.

\section{Effective field theory}
\label{sect:eft}

In heavy-baryon chiral perturbation theory ($\chi$PT), it is usual to 
define the Sachs electromagnetic form factors, $G_{E,M}$, which parametrize 
the matrix element 
 for the quark current, $J_\mu$, as 
\begin{align}
&\lb B(p')| J_\mu|B(p)\rk = \nn\\
\label{eqn:matelem}
&\bar{u}^{s'}(p')\left\{\ro{v}_\mu\, G_E(Q^2)
+\f{i\ep_{\mu\nu\rho\si}\ro{v}^\rho\,
 S_\ro{v}^\sigma\, q^\nu}{m_B}\,G_M(Q^2)\right\}u^s(p),
\end{align}
where $\ro{v}$ is the velocity of the baryon and
 $Q^2$ is the  positive momentum transfer $Q^2 = -q^2 = -(p'-p)^2$.
Lattice QCD results are often constructed from an alternative representation, 
using the form factors $F_1$ and $F_2$, which are
 called the Dirac and Pauli form factors, respectively.
The Sachs form factors are simply linear combinations of $F_1$ and $F_2$ 
\begin{align}
G_E(Q^2) &= F_1(Q^2) - \f{Q^2}{4m_B^2}F_2(Q^2)\,,\\
G_M(Q^2) &= F_1(Q^2) + F_2(Q^2)\,.
\end{align}
In the heavy-baryon formulation of Eq.~(\ref{eqn:matelem}), 
the spin operator,  
$S^\mu_\ro{v} = -\f{1}{8}\g_5[\g^\mu,\g^\nu]\ro{v}_\nu$, is used  
\cite{Jenkins:1990jv,Jenkins:1991ts}.
The momentum transfer dependence in the  
electric form factor, $G_E(Q^2)$, 
allows a charge 
radius 
to be defined in the usual manner 
\eqb
\label{eqn:raddefn}
\rad_E = \lim_{Q^2\rightarrow 0}-6\frac{\cd G_E(Q^2)}{\cd Q^2}.
\eqe

\subsection{Loop integral definitions}
\label{sect:int}

\begin{figure}
\centering
\includegraphics[height=100pt]{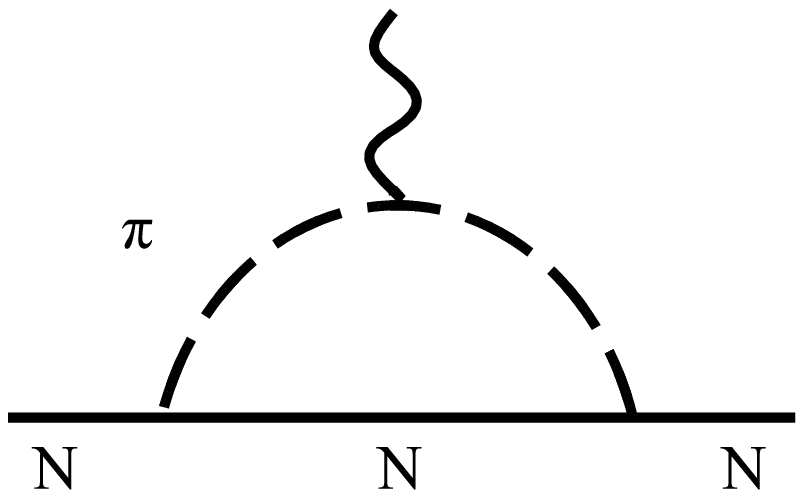}
\caption{\footnotesize{The pion loop contributions to the electric charge radius of a nucleon. All charge conserving pion-nucleon transitions are implicit.}}
\label{fig:emSEa}
\vspace{2mm}
\centering
\includegraphics[height=100pt,angle=0]{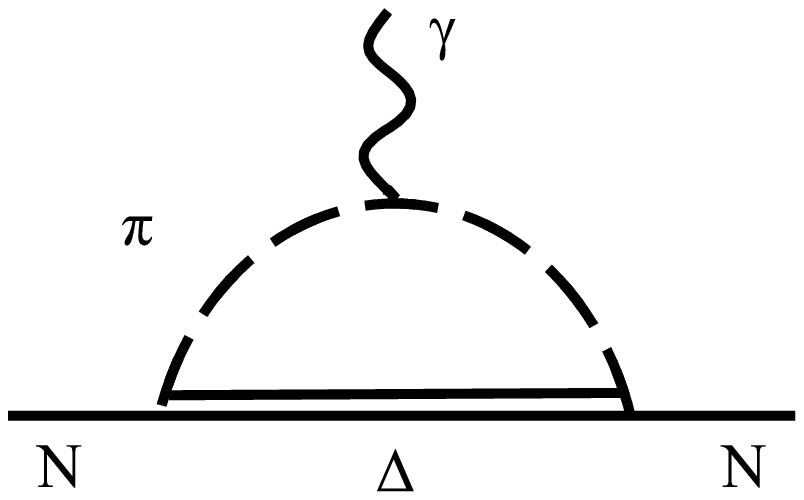}
\caption{\footnotesize{The pion loop contribution to the electric charge radius
 of a nucleon, allowing transitions to the nearby and strongly-coupled $\Delta$ baryons.}}
\label{fig:emSEb}
\vspace{4mm}
\centering
\includegraphics[height=100pt,angle=0]{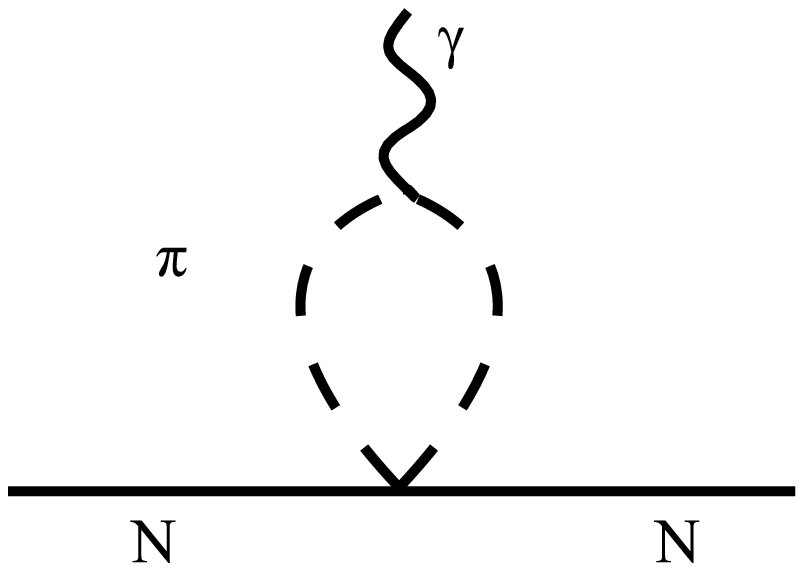}
\caption{\footnotesize{The tadpole contribution at $\ca{O}(m_q)$ to the 
electric charge radius of a nucleon.}}
\label{fig:emSEc}
\end{figure}

The loop integrals, in the continuum
limit, that contribute to the electric form factor of the nucleon 
are invariant under arbitrary translations of the internal momentum 
$\vec{k}\rightarrow \vec{k}+\de \vec{k}$.  
However, a finite-volume sum over discrete loop momenta 
is only invariant if $\de \vec{k}$ 
is an allowed value of momentum on the lattice. 
The loop integrals in the 
heavy-baryon approximation 
that 
correspond to the leading-order diagrams in 
Figs.~\ref{fig:emSEa} through \ref{fig:emSEc}
are obtained by 
performing the pole integration for $k_0$  
\begin{align}
\label{eqn:SiL}
\ca{T}_{N}(\vec{q}^2) &= 
\frac{\chi_{N}}{3\pi}\!\int\!\!\mathrm{d}^3 k 
\frac{(k^2-\vk\cdot\vq)}
{\omega_{\vk}\omega_{\vk-\vq}(\omega_{\vk}
+\omega_{\vk-\vq})},\\
\label{eqn:SiLD}
\ca{T}_{\De}(\vec{q}^2) &=  
\frac{\chi_\De}{3\pi}\!\int\!\!\mathrm{d}^3 k 
\frac{(k^2-\vk\cdot\vq)}
{(\omega_{\vk} + \Delta)(\omega_{\vk-\vq} + \Delta)(\omega_{\vk}
+\omega_{\vk-\vq})},\\
\ca{T}_{\ro{tad}}(\vec{q}^2) &=  \frac{\chi_{t}}{\pi}\!\int\!\!\mathrm{d}^3 
k \frac{1}
{\omega_{\vk}+\omega_{\vk-\vq}},
\label{eqn:SiT}
\end{align}
where $\om_{\vec{k}} = \sqrt{{\vec{k}}^2 + m^2_\pi}$, $m_\pi$ 
is the pion mass, 
and $\De$ is the Delta-nucleon mass-splitting. Note that 
each integral does not explicitly depend on $Q^2$, but depends on 
the three-momentum transfer squared, $\vec{q}^2$. 
The chiral coefficients $\chi_N$, $\chi_\De$ and $\chi_t$ are 
derived from couplings arising in the Lagrangian of chiral perturbation 
theory   
\cite{Wang:2008vb}
\begin{align}
\chi_N^{\ro{prot}} &= \f{5}{16\pi^2f_\pi^2}(D+F)^2 = -\chi_N^{\ro{neut}},\\
\chi_\De^{\ro{prot}} &= -\f{5}{16\pi^2f_\pi^2}\f{4 \,\mathcal{C}^2}{9} = 
-\chi_\De^{\ro{neut}},\\
\chi_t^{\ro{prot}} &= -\f{1}{16\pi^2f_\pi^2} = -\chi_t^{\ro{neut}},
\end{align}
where the value of the pion decay constant is $f_\pi = 92.4$ MeV. 
The values for the couplings are estimated from the 
 $\SU(6)$
 flavor-symmetry relations \cite{Jenkins:1991ts,Lebed:1994ga} 
and from phenomenology: $D = 0.76$, $F = \f{2}{3}D$
and $\ca{C} = -2D$. 

To obtain the integrals that 
 allow the determination of the quark mass expansion of
 the electric charge radius, 
one takes the derivative of each infinite-volume integral, 
$\ca{T}(\vec{q}^2)$, 
with respect to momentum transfer, $\vec{q}^2$, as $\vec{q}^2\rightarrow 0$: 
\eqb
T = \lim_{\vec{q}^2\rightarrow 0}-6\f{\cd \ca{T}(\vec{q}^2)}{\cd \vec{q}^2}.
\label{eqn:siSi}
\eqe 
This derivative is equivalent to that of Eq.~(\ref{eqn:raddefn}) 
in the Breit frame, defined by zero energy transfer to the nucleon 
($q^\mu = (0,\vec{q})$). Using the derivative forms, $T$, therefore 
allows one to recover the familiar chiral expressions for the 
quark mass dependence of the charge radii. 

\subsection{Finite-volume corrections}
\label{sect:fvc}

In the analysis of finite-volume effects within $\chi$EFT, one requires 
an evaluation of the correction associated with
 replacing the continuum loop integrals 
by finite sums. This correction is expressed in the form 
\eqb
\label{eqn:sum}
 \delta_L[\ca{T}] = \chi\left[\f{{(2\pi)}^3}{L_x L_y L_z}\sum_{k_x,k_y,k_z} - 
\int\!\mathrm{d}^3k\right]\ca{I}\,,
\eqe
for an integrand, $\ca{I}$.  
This is not so straightforward in the case of the charge radius, 
 which involves a $\vec{q}^2$ derivative. 
Because of the fact that only certain, 
discrete values of momenta are allowed on the lattice,   
only a finite-difference equation may 
be constructed from these allowed momenta. 
The finite-difference equation, ideally, would be constructed from 
the lowest value of $\vec{q}^2$ available on the lattice,   
 $\vec{k}_{\ro{min}}^2 = (2\pi\vec{n}/L)^2$ 
(where $\vec{n}$ is a lattice unit vector). 
 This is not possible to do in the Breit frame, 
where the lowest $\vec{q}^2$ value is at least $2(2\pi\vec{n}/L)^2$, 
which, on a moderate lattice size of $3$ fm, is approximately $(0.58$ GeV$)^2$. 
In order to obtain a suitable estimate of the slope of the form factor 
at $\vec{q}^2 = 0$, 
 a procedure is outlined for evaluating finite-volume 
corrections using the lowest available $\vec{q}^2$ value. 
Since the finite-volume corrections are applied directly to the form factor,
 an infinite-volume radius may be estimated. 
Thus, the quark mass behaviour of the radius may be examined 
independently of the finite-volume effects.

\begin{figure}
\centering
\includegraphics[height=0.8\hsize, angle = 90]{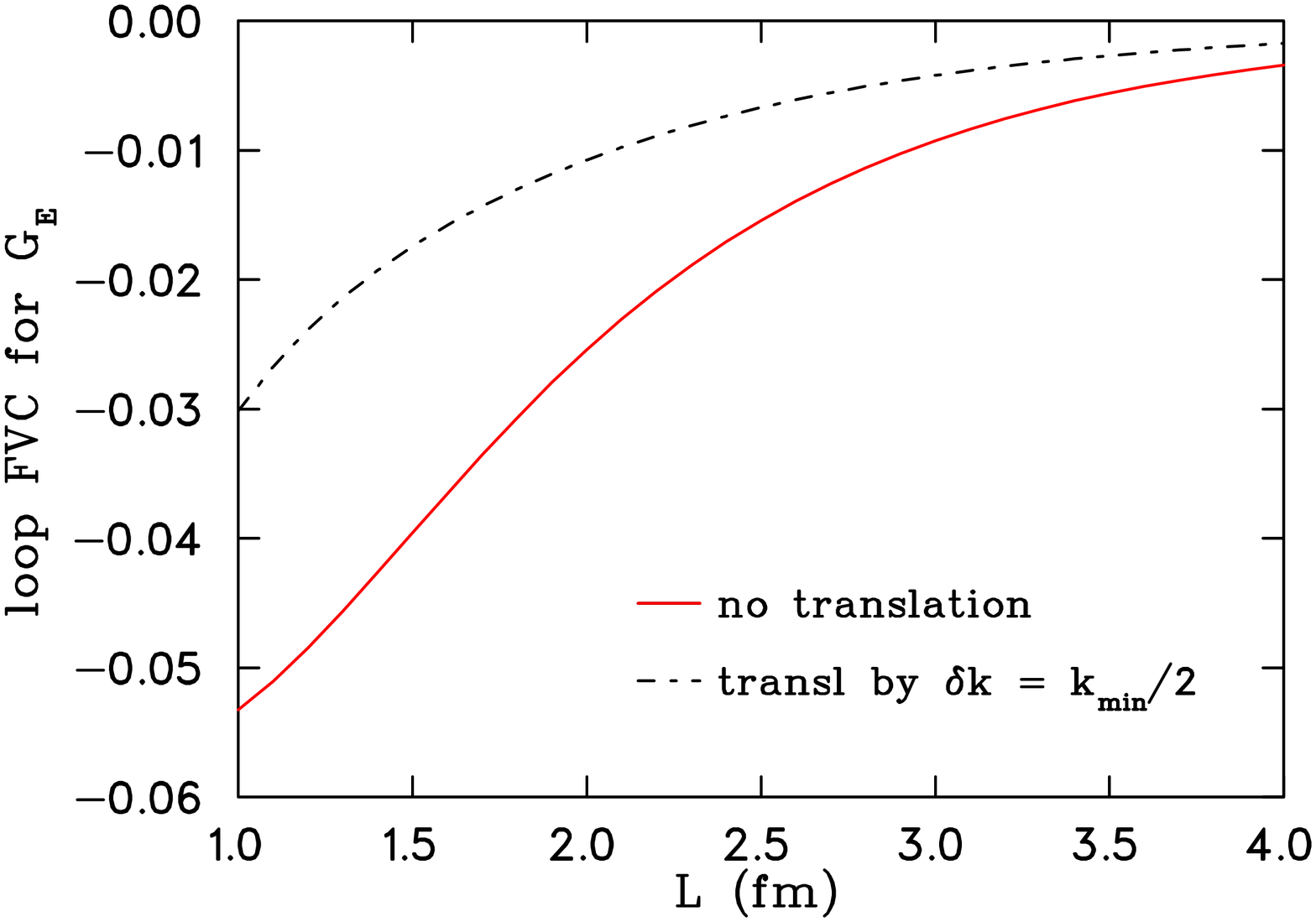}
\caption{\footnotesize{(color online). Finite-volume correction  for the leading-order loop integrals contributing to $G_E$, with $\vec{q} = \vec{k}_{\ro{min}}$. The choice of $\de \vec{k} = \vec{k}_{\ro{min}}/2$ is not an allowed value on the lattice. The momentum translated and untranslated behaviour of the finite-volume correction are inconsistent with each other.}}
\label{fig:loopfvcQsq}
\vspace{7mm}
\includegraphics[height=0.8\hsize, angle = 90]{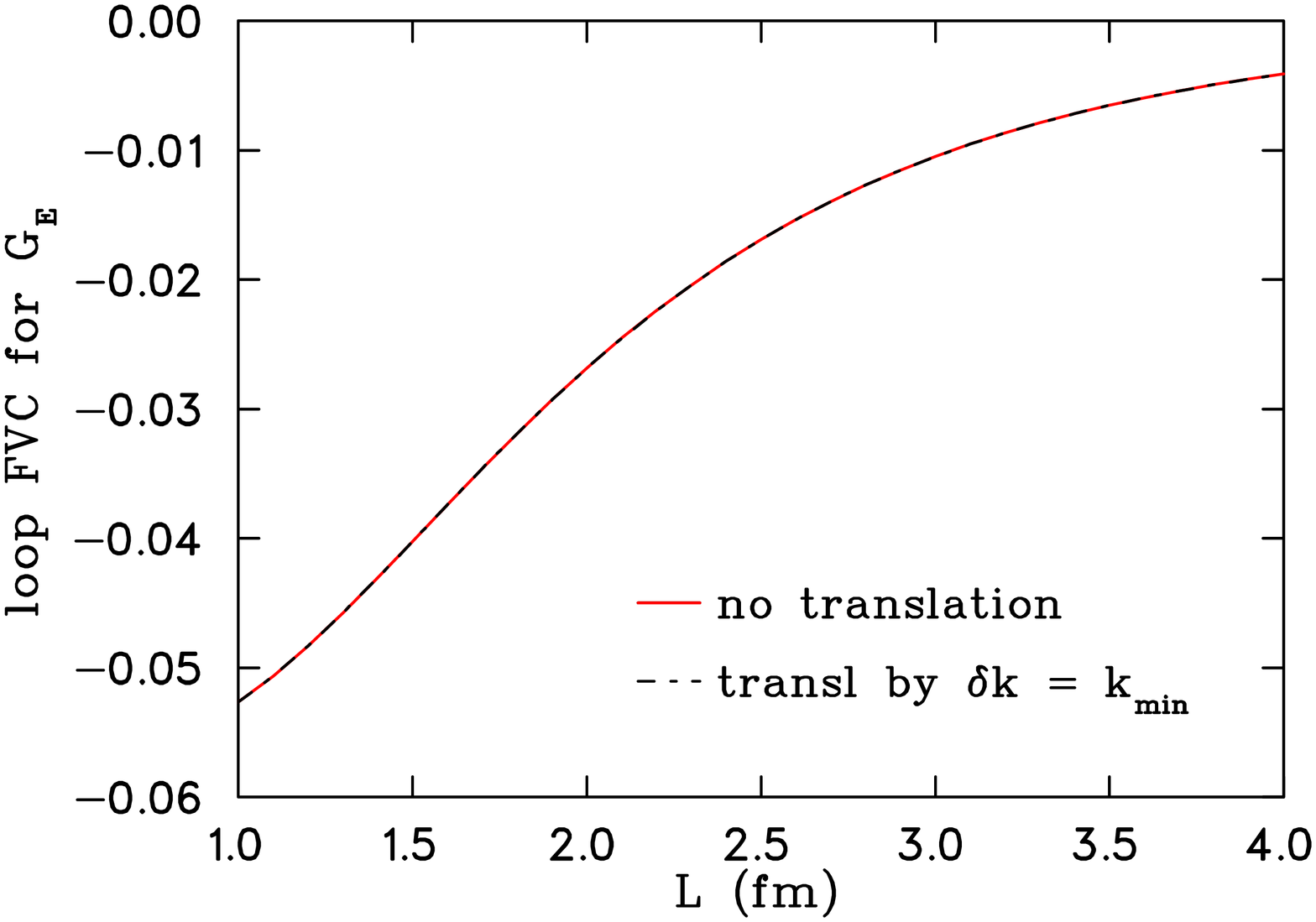}
\caption{\footnotesize{(color online). Finite-volume correction for the leading-order loop integrals contributing to $G_E$, with $\vec{q} = \vec{k}_{\ro{min}}$. The choice of $\de \vec{k} = \vec{k}_{\ro{min}}$ is an allowed value on the lattice. Therefore, the momentum translated and untranslated behaviour of the finite-volume correction are identical.}}
\label{fig:loopfvc4Qsq}
\end{figure}

In order to illustrate the effect of using 
the loop integrals evaluated at allowed, and unallowed, values of momentum 
transfer, $\vec{q}$, on the lattice, a comparison is shown in Figs.~\ref{fig:loopfvcQsq} and \ref{fig:loopfvc4Qsq} 
in which the finite-volume correction to the one-pion loop integral 
(Eqs.~(\ref{eqn:SiL}) \& (\ref{eqn:SiLD}))   
is plotted as a function of box size, $L$. In Fig.~\ref{fig:loopfvcQsq}, %4, 
the momentum 
transfer, $\vec{q}$, is taken to be 
$\vec{k}_{\ro{min}}$. In evaluating the loop sums, 
a momentum translation of $\delta \vec{k} = \vec{k}_{\ro{min}}/2$ 
is not an allowed value, and the finite-volume correction 
is inconsistent. This is a consequence of spoiling the continuous 
symmetry by the discretization of the momenta. 
In Fig.~\ref{fig:loopfvc4Qsq}, 
$\delta \vec{k} = \vec{k}_{\ro{min}}$ is an allowed value, 
and the translated and untranslated 
results for the finite-volume correction are identical.

Note that it is possible to obtain a
 momentum transfer that is not a standard lattice vector  
 by introducing twisted boundary conditions on 
 the valence quark propagators. In tuning the twist angle to access 
non-integer momentum states, one must be 
careful to account for modified contributions to finite-volume corrections, 
as discussed in Ref.~\cite{Jiang:2008ja}.

\section{Procedure for obtaining charge radii at finite volume}
\label{sect:procedure}

In extracting radii from lattice simulations, 
one should start with the form factors as extracted from the lattice, 
and only convert them into radii (using a suitable Ansatz for the 
$Q^2$-behaviour) 
after correcting for 
lattice finite-volume effects.  
The finite-volume correction to $G_E(Q^2)$ is achieved by subtracting
the difference between the sum and integral loop contributions, from 
Eq.~(\ref{eqn:sum}) 
\eqb
G_E(Q^2) = G_E^L(Q^2) - \de_L[\ca{T} (\vec{q}^2)],
\eqe
where $\ca{T} = \ca{T}_N + \ca{T}_\De + \ca{T}_\ro{tad}$. 
In order to maintain conservation of charge, a subtraction 
 of 
 the finite-volume 
corrections at zero momentum transfer, $\de[\ca{T}(\vec{q}^2=0)]$, 
is introduced  
\eqb
\label{eqn:fvc}
 \delta_L[\ca{T}(\vec{q}^2)] \rightarrow
\delta_L[\ca{T}(\vec{q}^2)] - \delta_L[\ca{T}(0)]. 
\eqe
The
second term of Eq.~(\ref{eqn:fvc}) ensures that both infinite- and 
finite-volume form 
factors are correctly normalized,
i.e.  
$G_E^{L}(0) = 1 = G_E(0)$, independent of the nucleon momentum. 
This normalization procedure assumes that charge conservation is 
satisfied in a finite volume, as demonstrated by the following 
numerical analysis. Within the framework of $\chi$EFT, preservation of the 
lattice Ward-Takahashi Identity in a finite volume has been addressed in 
Ref.~\cite{Hu:2007eb}. 

A lattice QCD calculation was undertaken to demonstrate charge conservation in the rest frame, and in several boosted frames of increasing momenta. Using the temporal component of a conserved vector current, and the prescription outlined in Ref.~\cite{Boinepalli:2006xd}, the Sachs electric form factor, $G^{L}_{E}$, at zero momentum transfer, was calculated for external momenta: $\vec{p}$ = $\vec{0}$, $\vec{k}_\ro{min}$, $2 \, \vec{k}_\ro{min}$ and $3 \, \vec{k}_\ro{min}$, where $\vec{k}_\ro{min}$ is the minimum available three-momentum on the lattice. In all cases considered, the extracted value of $G^{L}_{E}(Q^2=0)$ was consistent with unity up to the level of precision present in the propagators used in the calculation, indicating that charge conservation is satisfied.

In general, matrix elements in lattice QCD depend on the external 
momentum, as discussed in detail in 
Refs.~\cite{Hu:2007eb,Jiang:2008ja}. 
That is, the breaking of SO$(4)$ symmetry entails a 
boost-dependence of the 
 current matrix elements in a finite volume.  
In that case, the frame, and current component, 
must be specified in obtaining finite-volume 
corrections to the matrix elements calculated on the lattice. 
However, in the heavy-baryon 
approximation, the finite-volume corrections to the leading-order 
loop contributions (Figs.~\ref{fig:emSEa} through \ref{fig:emSEc}) 
depend only on $\vec{q}^2$, and not on the boost of the initial or 
final nucleon state. Physically, this is simply a consequence of 
a lack of recoil energy-dependence in the intermediate nucleon 
propagator. This boost-invariance is also realized in Eq.~(A1) 
of Ref.~\cite{Jiang:2008ja}, by removing the twisted boundary condition 
and the partial quenching.

With the finite-volume corrected form factors at hand, 
the charge radii, $\rad_E$, can be recovered 
from the form factors by using a suitable $Q^2$ parametrization. 
At large $L$, 
and hence with numerous small $Q^2$ values, 
a formal $Q^2$-expansion from $\chi$EFT would be ideal, 
 such as that used in Ref.~\cite{Bauer:2012pv}. 
At smaller values of $L$, with limited $Q^2$ values, it is common 
to work with a more phenomenological Ansatz, such as the dipole form 
\eqb
G_E(Q^2_{\ro{min}}) = 
{\left(1 + \frac{Q^2_{\ro{min}}\rad_E}{12}\right)}^{-2}\,,
\eqe
where $Q^2_{\ro{min}} = \vec{k}^2_{\ro{min}} - (E_N' - E_N)^2$. 
Another example is  
an inverse quadratic with two 
fit parameters, as suggested by Kelly \cite{Kelly:2004hm} 
and used by Collins \textit{et al.} 
\cite{Collins:2011mk} 
\eqb
\label{eqn:quad}
G_E(Q^2) = \f{G_E(0)}{1 + \alpha Q^2 + \beta Q^4},
\eqe
with the charge radius obtained through $\rad_E = 6\,\alpha$. 
Once the  
infinite-volume charge radii have been obtained, 
 one then has best estimates for charge radii at any given set of 
lattice parameters. 

\begin{figure}
\centering
\includegraphics[height=0.8\hsize, angle = 90]{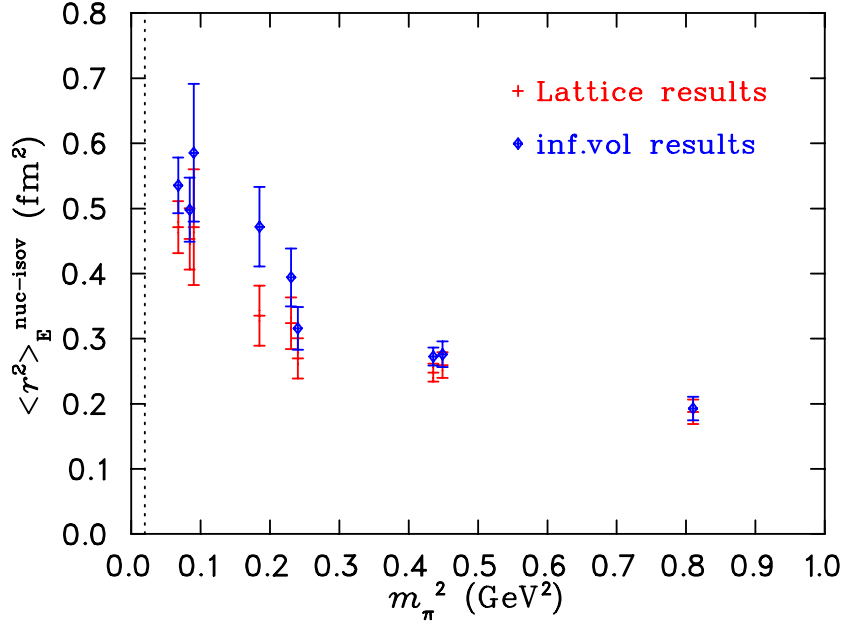}
\caption{\footnotesize{(color online). Lattice QCD results for the isovector nucleon electric charge radius from QCDSF Collaboration at their original box sizes, and corrected to infinite volume.}}
\label{fig:QCDSF}
\end{figure}

As a demonstration of the method, lattice QCD results
 from QCDSF Collaboration \cite{Collins:2011mk} 
for the isovector nucleon electric charge 
radius are corrected to 
infinite volume, and shown in Fig.~\ref{fig:QCDSF}. 
The lattice calculation  
uses the two-flavor $\ca{O}(a)$-improved
 Wilson quark action. 
%For each non-zero $Q^2$ value, the form factors are calculated by averaging 
%over all possible directions of $\vec{q}$, the directions of the 
%electromagnetic current and the polarizations of the nucleon,    
%at $\vec{p}'=0$. While this means that all current components 
%have been averaged, it is expected that %the statistical uncertainty in 
%the result is predominantly governed 
%by the time component. 
%Future studies should 
%nevertheless be cautious of the additional corrections associated 
%with the spatial components of the currents \cite{Jiang:2008ja}.
%
The points displayed satisfy $m_\pi L > 3$, and the box sizes for each point 
 range from $1.7$ to $2.9$ fm. 
The infinite volume form factors are calculated via
\begin{align}
G_E(Q^2) &= G_E^L(Q^2) - (\de_L[\ca{T}(\vec{q}^2)] - \de_L[\ca{T}(0)]), \\
\mbox{with}\,\, \ca{T}(\vec{q}^2) &= \ca{T}_N(\vec{q}^2) + \ca{T}_\De(\vec{q}^2) 
+ \ca{T}_\ro{tad}(\vec{q}^2).
\end{align}

In the construction of the charge radii, %following Ref.~\cite{Collins:2011mk}, 
the $Q^2$-extrapolation Ansatz 
of Eq.~(\ref{eqn:quad}) is used, to be consistent with 
Ref.~\cite{Collins:2011mk}.  
Fig.~\ref{fig:QCDSF} clearly shows that the infinite-volume points are  
larger in radius than the finite-volume points, and closer to the 
experimental value of $\langle r^2\rangle_E^\ro{isov} = 0.88$ fm$^2$ 
\cite{Mohr:2008fa,Nakamura:2010zzi}. 

The technique as described thus provides charge radii  
at different quark/pion masses, and hence enables the use of continuum 
$\chi$EFT to fit the quark mass dependence. Details of such an 
extrapolation will appear in a forthcoming paper. 

\section{Conclusion}
\label{sect:conc}
 
Direct finite-volume corrections to charge radii are not well-defined on the 
lattice. 
The use of continuous derivatives in constructing the electric 
charge radius 
leads to inconsistent results in the finite-volume corrections. 
Alternatively, a satisfactory definition of radii can be achieved by 
implementing 
finite-volume corrections 
to the electric form factors directly, 
 evaluated at discrete values of $\vec{q}^2$. 
Subsequently, the resultant finite-volume-corrected form factors  
may then be converted into charge radii using an appropriate  
extrapolation in $Q^2$. 
A suitable definition of charge radius for comparison with 
experiment and continuum $\chi$EFT analysis has thus been obtained.

We would like to thank Brian Tiburzi for helpful discussions. 
This research 
is supported by the Australian Research Council through grant DP110101265.

%\section*{References}
%BIBLIOGRAPHY
%\bibliographystyle{model3-num-names}
\bibliographystyle{apsrev}
\bibliography{fvc}

\end{document}